# Flexible on-chip mode-division switching with a new mode converter design

*Wanyuan Liu*
*email: wylcq8@gmail.com*

**There have been some significant advances in optical-fiber multimode communications. And important effort in on-chip communication has also been recently reported for realizing mode-division multiplexing (MDM). Because of the difficulty in lack of mode converter which can freely perform mode converting function among different waveguide modes, flexible on-chip mode-division switching communication has been rarely reported. Here we show the first demonstration of mode converter which can convert an arbitrary waveguide mode into any other modes with the help of phase tuning. With that mode converter, two flexible on-chip mode-division switching architectures are proposed. That architecture is compatible with wavelength-division multiplexing (WDM) and can enable further scaling of the on-chip communication bandwidth. Our demonstration suggests great potential of mode-divisional switching in the vision of fully flexible, dense, on-chip optical networks.**

The data switching is an important function for flexible and advanced optical networks and considerable research has been given to wavelength division multiplexed optical-fiber systems. In these systems, data exchange, which is also known as wavelength exchange or interchange in the wavelength domain, is an important technique which can efficiently make use of network resources and improve network performance. The switchable and flexible function is also significant for advanced and reconfigurable photonic-integrated optical network because of data routing between the processor and memory systems on a chip multiprocessor. However, because of the expensive cost and complicated process requirement, it is hard to realize the multiple laser sources with different wavelengths for on-chip optical interconnection applications. Space-division multiplexing (SDM), which only employs a single wavelength carrier, provides a new dimension to meet the growing demand on transmission capacity of our communication systems, by utilizing the spatial modes to carry multiple optical signals simultaneously. Significant effort in optical-fiber multimode communications has been put in recent years into realizing mode-division multiplexing. Numerous advances on multimode communication in fibers have so far been reported, such as space-division multiplexing in multi-core fibers[1-5] and mode-division multiplexing in few-mode fibers[6-14]. Fiber communications are also trying to combine multimode operation with other multiplexing technologies, such as WDM[15], to further scale the communication bandwidth transmitted per fiber. For the demand of on-chip and off-chip high-bandwidth-density interconnects, photonic-integrated mode division multiplexing attracts more and more attentions by taking advantages of increasing the transmission capacity efficiently. Several designs for on-chip mode multiplexers have been proposed previously, including designs based on Mach-Zehnder interferometers[16,17], multimode interference couplers[18-20], asymmetric directional couplers[21-23] and y-junctions[24-26]. Previous implementations of mode-division multiplexing switch[27] and mode exchange using tapered directional coupler[28] have also been demonstrated. However, those proposed devices, no matter mode multiplexers, multimode switches or mode exchange schemes, suffer from several drawbacks, including a limited number of optical modes or a lack of selectivity among user's wanted modes, which are not suited to fully flexible, dense, on-chip optical networks. Some of the key challenges of realizing on-chip MDM-enable flexible interconnects lie in creating the mode converter and (de)multiplexer which can selectively deal with arbitrary mode. But it is really hard to finish those designs, because the mode confinements in a multimode waveguide vary significantly between the different modes. That is why on-chip data switching based on MDM is rarely considered, while it is highly desirable to increase robustness and throughput of the chip-level network utilizing mode multiplexing.

## Mode converter

Here, we present a new mode converter design, as shown in Figure 1a, which accomplishes the mode conversion function from fundamental mode to any other optical modes and provides wide bandwidth operation. It comprises a middle multimode transport waveguide consisting of several sections with tapering widths from a single mode waveguide to different multimode waveguides, and two single mode waveguides extending along both sides of the middle multimode waveguide. To show a mode conversion function from TE0 to TE1 and TE0 to TE2, that middle multimode waveguide tapers widths ranging from 0.45 um to 1.41 um. When the multimode waveguide width corresponds to 450 nm, 930nm or 1.41 um, the effective indices of TE0, TE1 or TE2 modes, respectively, match the effective index of the TE0 mode of the 0.45 um waveguide[29]. The separation gaps between the upper lower single mode waveguide and multimode waveguide at TE0, TE1 and TE2 coupling regions are 250 nm, 200 nm and 200 nm, respectively. In that middle multi-mode waveguide, there is a fixed phase difference between the odd modes which are respectively excited by the upper and lower single mode waveguide with their fundamental mode inputting, and so does the even modes. Several phase-shifting elements are added to the upper single mode waveguide section to accomplish each mode conversion function. When single mode light is input into left side of the middle waveguide, it will be first split into two beams propagating in the upper and lower waveguides. And by tuning their phase difference, destructive interference or constructive interference can be controlled to perform the mode conversion function. Once mode conversion is accomplished, the converted mode will propagate steadily in the middle multimode waveguide.

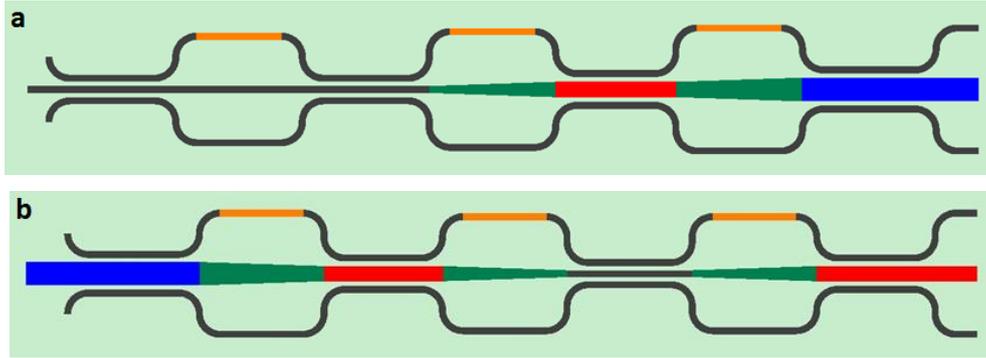

**Figure 1 | The schematics of the proposed photonic integrated mode converters. a**, The schematic diagram of the mode converter performing conversion function from TE0 to the other arbitrary optical mode. **b**, The schematic diagram of the mode converter performing conversion function from arbitrary optical mode to any other mode.

A further improved design is shown in Figure 1b. That structure accomplishes the mode conversion function from an arbitrary optical mode to the other mode, by converting the arbitrary mode first into TE0 mode which is then be converted to any other wanted mode. That mode conversion function, like the wavelength conversion function to WDM, is the key technology to mode-division multiplexing.

For the convenience of simulation, here the width of the upper single mode waveguide was changed to realize the phase tuning. And the mode conversion from TE0 mode to the TE0, TE1, or TE2 modes and the conversion from TE2 mode to TE1 mode are taken as the examples. Based on 3D FDTD method, the simulated electric field distribution of mode conversion function among those different modes with our mode converter is shown in Figure 2.

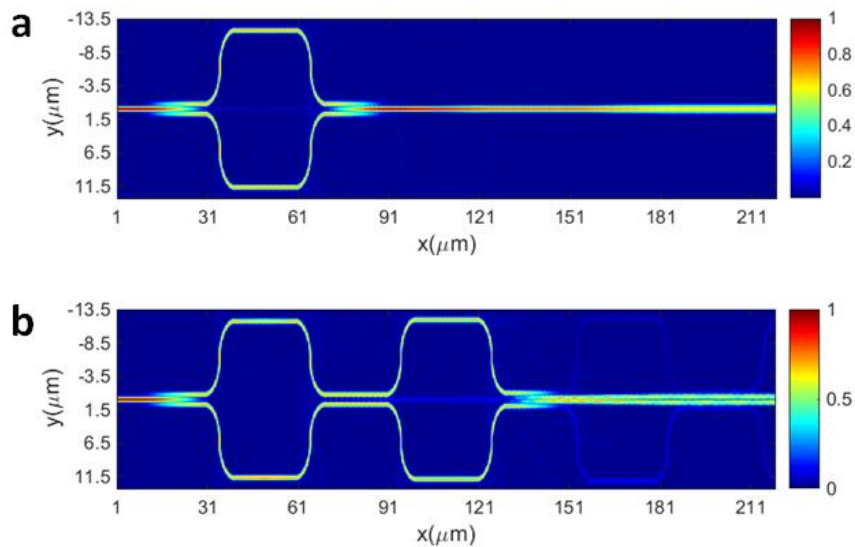

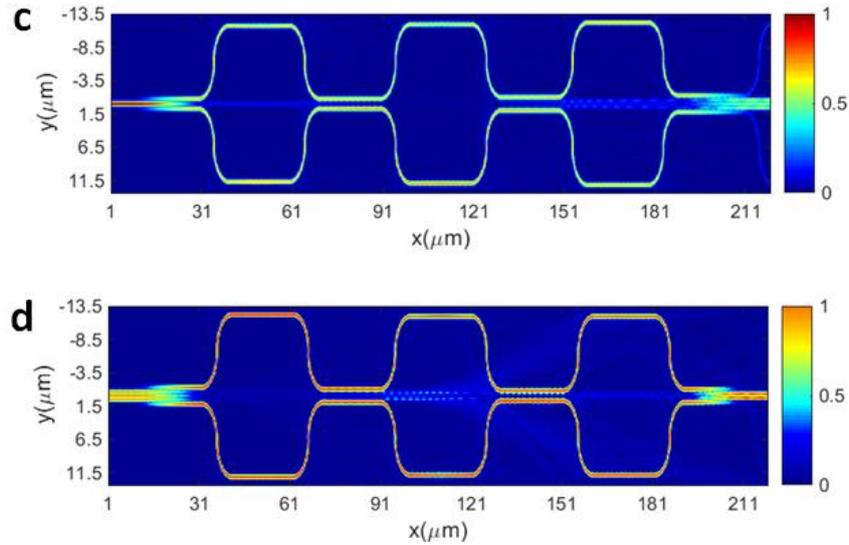

**Figure 2 | Electric field distribution of mode conversion among different modes. a**, Mode conversion from TE0 to TE0. **b**, Mode conversion from TE0 to TE1. **c**, Mode conversion from TE0 to TE2. **d**, Mode conversion from TE2 to TE1.

The power of converted mode from different modes can be computed with 3D FDTD and depicted in Fig. 3. It shows the transmission spectrum of mode conversion among different modes. The simulated insertion loss of the conversion from TE0 to TE0 at λ=1,550nm is about 0.2 dB. The insertion loss of mode conversion from TE0 to TE1 at λ=1,550nm is about 0.2 dB. And the insertion loss from TE0 to TE2 at λ=1,550nm is 0.5 dB. The insertion loss from TE2 to TE1 at λ=1,550nm is about 0.6 dB. All of those mentioned mode conversion function performs a broadband performance in C-band. There are two main things contributed to the insertion loss. One is the growing amount of bends, tapers and the coupling times, and another is the imprecise phase tuning. That insertion loss can be reduced by increasing the radius of bend, extending the length of taper and further optimizing on the waveguide coupling region. To realize a precise phase tuning, electro-optic effect or thermo-optic effect can selected for users in actual silicon chip fabrication.

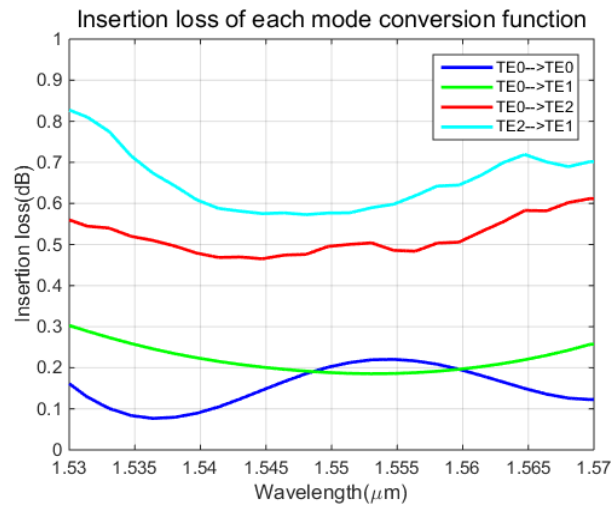

**Figure 3 | Insertion loss of mode conversion among different modes.**

Because of the strong refractive index tuning capacity of thermo-optic effect[30], it can be used to control the phase difference between those two single mode arms. To avoid absorption, the optical mode should be well separated from the heater, and should also be close enough to couple heat into the waveguide. Some configuration parameters of the waveguide and heaters can refer to the published paper[31]. The single mode waveguide in this paper is 450 nm wide by 250 nm thick.

## Mode-division switching architecture

Based on the mode converter shown in this paper, here we present two new communication architecture designs which can accomplish the flexible mode-division switching function. The first design, shown in Figure 4a, comprises the mode converter with mode (de)multiplexers[23] connected at its input and output port. With the help of mode converting function from that converter, each inputting light from the left side can be switched to any of the outputting port on the right. That is a simple design based on the mode converter presented in this paper, but its disadvantage is that light at the left side cannot be injected at the same time. Because the mode converter can only perform the mode converting function on one waveguide mode at one time, it cannot work on different modes simultaneously.

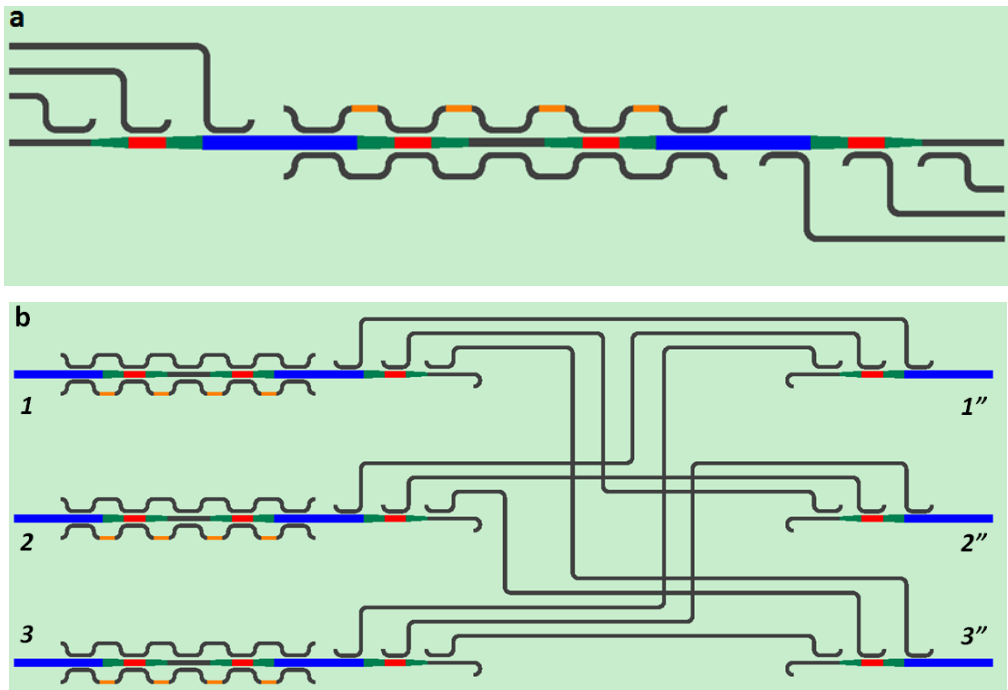

**Figure 4 | The schematics of the proposed mode-division switching architectures. a**, The schematic diagram of the simple architecture design used for mode-division switching. **b**, The schematic diagram of the improved architecture design used for mode-division switching.

The improved design is shown in Figure 4b. There is a mode converter connected at each input port in this design. And a mode de-multiplexer is connected with that each mode converter. Light from that de-multiplexer can be coupled into the output port on the right with the help of another multiplexer. Because of the mode converting function of the mode converter, each inputting light from the left side can be switched into any of the outputting port on the right. The problem that light at the left side cannot be injected at the same time in the first design is solved in this improved design. And the most important is that the simultaneously different inputting light from the left side can be switched to output from the same port or each to different output ports on the right. So that is a non-blocking and flexible switching architecture and there is obviously improvement on its communication capacity compared to the common reported silicon switch matrix with the same scale.

Here we extract the S parameters which are from the 3D FDTD simulation results of those basic elements included in the above mentioned architecture to show the propagating performance of the second design in Figure 4b. The propagation that TE0 mode input into the first port on the left side and then TE0 output from the second port on the right is taken as the example. In that propagation, it is necessary to convert the input TE0 mode first into TE1 mode, and then it can be coupled into the second port on the right. The transmission spectrum is shown in Figure 5.

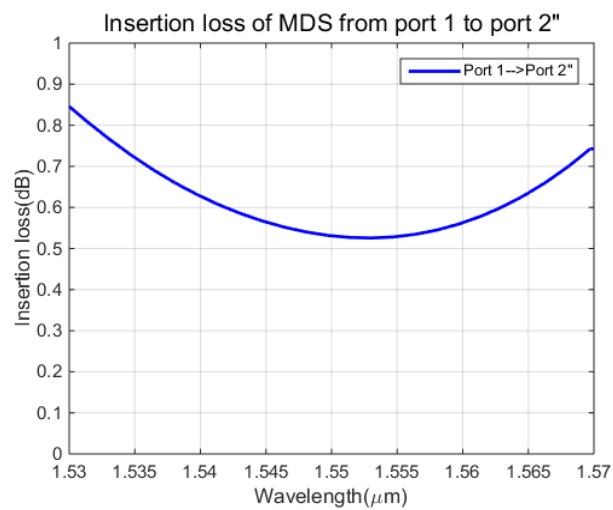

**Figure 5 | Insertion loss of mode Mode-division switching from port 1 to port 2".**

In summary, a novel mode converter design which can convert an arbitrary waveguide mode into any other mode with the help of phase tuning have been first demonstrated in this paper. That free mode conversion function is the key technology for mode-division multiplexing. In particular, we present two new communication architecture designs which can accomplish the flexible mode-division switching function. Our work represents an important step towards the realization of a practical on-chip mode-division multiplexing application. For its compatibility with WDM, it can be combined with wavelength-divisional photonic switching to further improve the performance of integrated optical networks.

# References

1. Sakaguchi, J. et al. 109-Tb/s (7x97x172-Gb/s SDM/WDM/PDM) QPSK transmission through 16.8-km homogeneous multi-core fiber, in Optical Fiber Communication Conference (Optical Society of America) (2011).
2. Feuer, M. D. et al. ROADM System for Space Division Multiplexing with Spatial Superchannels, in Optical Fiber Communication Conference (Optical Society of America) (2013).
3. Sakaguchi, J. et al. 19-core fiber transmission of 19x100x172-Gb/s SDM-WDM-PDM-QPSK signals at 305Tb/s, in Optical Fiber Communication Conference (Optical Society of America) (2012).
4. Takara, H. et al. 1.01-Pb/s (12 SDM/222 WDM/456 Gb/s) Crosstalk-managed Transmission with 91.4-b/s/Hz Aggregate Spectral Efficiency, in European Conference and Exhibition on Optical Communication (Optical Society of America) (2012).
5. Qian, D. et al. 1.05Pb/s Transmission with 109b/s/Hz Spectral Efficiency using Hybrid Single- and Few-Mode Cores, in Frontiers in Optics (Optical Society of America) (2012).
6. Randel, S. et al. 6×56-Gb/s mode-division multiplexed transmission over 33-km few-mode fiber enabled by 6×6 MIMO equalization. Opt. Express 19, 16697-16707 (2011).
7. Hanzawa, N. et al. Demonstration of mode-division multiplexing transmission over 10 km two-mode fiber with mode coupler, in Optical Fiber Communication Conference and Exposition (OFC/NFOEC) (2011).
8. Ryf, R. et al. Mode-Division Multiplexing Over 96 km of Few-Mode Fiber Using Coherent 6 × 6 MIMO Processing. J. Lightwave Technol. 30, 521-531 (2012).
9. Salsi, M. et al. Mode-Division Multiplexing of 2 x 100 Gb/s Channels Using an LCOS-Based Spatial Modulator. Lightwave Technology, Journal of 30, 618-623 (2012).
10. Al Amin, A. et al. Dual-LP11 mode 4x4 MIMO-OFDM transmission over a two-mode fiber. Opt. Express 19, 16672-16679 (2011).
11. Bai, N. et al. Mode-division multiplexed transmission with inline few-mode fiber amplifier. Opt. Express 20, 2668-2680 (2012).
12. Ryf, R. et al. 32-bit/s/Hz Spectral Efficiency WDM Transmission over 177-km Few-Mode Fiber, in Optical Fiber Communication Conference (Optical Society of America) (2013).
13. Ip, E. et al. 146λx6x19-Gbaud Wavelength- and Mode-Division Multiplexed Transmission over 10x50-km Spans of Few-Mode Fiber with a Gain-Equalized Few-Mode EDFA, in Optical Fiber Communication Conference (Optical Society of America) (2013).
14. Li, A., Al Amin, A., Chen, X. & Shieh, W. Transmission of 107-Gb/s mode and

polarization multiplexed CO-OFDM signal over a two-mode fiber. Opt. Express 19, 8808-8814 (2011).
15. Richardson, D. J., Fini, J. M. & Nelson, L. E. Space-division multiplexing in optical fibres. Nat Photon 7, 354-362 (2013).
16. Bagheri, S. & Green, W. in 6th IEEE International Conference on Group IVPhotonics, 166–168 (2009).
17. Huang, Y., Xu, G. & Ho, S.-T. An ultracompact optical mode order converter. Phot. Technol. Lett. 18, 2281–2283 (2006).
18. Kawaguchi, Y. & Tsutsumi, K. Mode multiplexing and demultiplexing devices using multimode interference couplers. Electron Lett. 38, 1701–1702 (2002).
19. Uematsu, T., Ishizaka, Y., Kawaguchi, Y., Saitoh, K. & Koshiba, M. Design of a compact two-mode multi/demultiplexer consisting of multimode interference waveguides and a wavelength-insensitive phase shifter for mode-division multiplexing transmission. J. Lightw. Technol. 30, 2421–2426 (2012).
20. Leuthold, J., Hess, R., Eckner, J., Besse, P. A. & Melchior, H. Spatial mode filters realized with multimode interference couplers. Opt. Lett. 21, 836–838 (1996).
21. Greenberg, M. & Orenstein, M. Multimode add-drop multiplexing by adiabatic linearly tapered coupling. Opt. Express. 13, 9381–9387 (2005).
22. Ding, Y. et al. On-chip two-mode division multiplexing using tapered directional coupler-based mode multiplexer and demultiplexer. Opt. Express. 21, 10376–10382 (2013).
23. Dai, D., Wang, J. & Shi, Y. Silicon mode (de) multiplexer enabling high capacity photonic networks-on-chip with a single-wavelength-carrier light. Opt. Lett. 38, 1422–1424 (2013).
24. Lee, B.-T. & Shin, S.-Y. Mode-order converter in a multimode waveguide. Opt. Lett. 28, 1660–1662 (2003).
25. Love, J. D. & Riesen, N. Single-, few-, and multimode y-junctions. J. Lightw Technol 30, 304–309 (2012).
26. Driscoll, J. B. et al. Asymmetric y junctions in silicon waveguides for on-chip mode-division multiplexing. Opt. Lett. 38, 1854–1856 (2013).
27. BRIAN STERN. et al. On-chip mode-division multiplexing switch. Optica 2(6), 530–535 (2015).
28. Z. Zhang. et al. On-chip optical mode exchange using tapered directional coupler. Sci. Rep. 5, 16072 (2015).
29. L.-W. Luo. et al. WDM-compatible mode-division multiplexing on a silicon chip. Nat. Commun. 5, 3069 (2014).
30. W. M. J. Green. et al. Ultra-compact reconfigurable silicon optical devices using micron-scale localized thermal heating. OFC/NFOEC 2007. (2007), pp. 1-3.
31. N. Sherwood-Droz. et al. Optical 4x4 hitless silicon router for optical Networks-on-Chip (NoC). Opt. Express 16(20), 15915–15922 (2008).